\documentclass[twocolumn]{aastex6}

\usepackage{color}

\newcommand{\ky}{K2-138}

\slugcomment{To Be Submitted to AJ}

\shorttitle{The K2-138 System}
\shortauthors{Christiansen et al.}

\begin{document}

\title{The K2-138 System: A Near-Resonant Chain of Five Sub-Neptune Planets Discovered by Citizen Scientists}

\author{Jessie~L.~Christiansen\altaffilmark{1,2}}
\author{Ian~J.~M.~Crossfield\altaffilmark{3,4}}
\author{Geert~Barentsen\altaffilmark{5,6}}
\author{Chris~J.~Lintott\altaffilmark{7}}
\author{Thomas~Barclay\altaffilmark{8,9}}
\author{Brooke~.D.~Simmons\altaffilmark{10,11}}
\author{Erik Petigura\altaffilmark{12}}
\author{Joshua E. Schlieder\altaffilmark{13}}
\author{Courtney D. Dressing\altaffilmark{4,10}}
\author{Andrew Vanderburg\altaffilmark{14}}
\author{David R. Ciardi\altaffilmark{1}}
\author{Campbell Allen\altaffilmark{7}}
\author{Adam McMaster\altaffilmark{7}}
\author{Grant Miller\altaffilmark{7}}
\author{Martin Veldthuis\altaffilmark{7}}
\author{Sarah Allen\altaffilmark{15}}
\author{Zach Wolfenbarger\altaffilmark{15}}
\author{Brian Cox\altaffilmark{16}}
\author{Julia Zemiro\altaffilmark{17}}
\author{Andrew W. Howard\altaffilmark{18}}
\author{John Livingston\altaffilmark{19}}
\author{Evan Sinukoff\altaffilmark{17,20}}
\author{Timothy Catron\altaffilmark{21}}
\author{Andrew Grey\altaffilmark{22}}
\author{Joshua J. E. Kusch\altaffilmark{22}}
\author{Ivan Terentev\altaffilmark{22}}
\author{Martin Vales\altaffilmark{22}}
\author{Martti H. Kristiansen\altaffilmark{23}}

\altaffiltext{1}{Caltech/IPAC-NASA Exoplanet Science Institute, M/S 100-22, 770 S. Wilson Ave, Pasadena, CA 91106 USA}
\altaffiltext{2}{jessie.christiansen@caltech.edu}
\altaffiltext{3}{Department of Astronomy \& Astrophysics, University of California, Santa Cruz, CA, USA}
\altaffiltext{4}{Sagan Fellow}
\altaffiltext{5}{NASA Ames Research Center, Moffett Field, CA 94035, USA}
\altaffiltext{6}{Bay Area Environmental Research Inst., 625 2nd St. Ste 209 Petaluma, CA 94952, USA}
\altaffiltext{7}{Department of Physics, University of Oxford, Denys Wilkinson Building, Keble Road, Oxford, OX1 3RH, UK}
\altaffiltext{8}{NASA Goddard Space Flight Center, 8800 Greenbelt Rd, Greenbelt, MD 20771, USA} 
\altaffiltext{9}{University of Maryland, Baltimore County, 1000 Hilltop Cir, Baltimore, MD 21250, USA}
\altaffiltext{10}{Einstein Fellow}
\altaffiltext{11}{Center for Astrophysics and Space Sciences (CASS), Department of Physics, University of California, San Diego, CA 92093, USA}
\altaffiltext{12}{Geological and Planetary Sciences, California Institute of Technology, Pasadena, CA, USA}
\altaffiltext{13}{NASA Goddard Space Flight Center, 8800 Greenbelt Rd, Greenbelt, MD 20771, USA}
\altaffiltext{14}{Harvard-Smithsonian Center for Astrophysics, 60 Garden Street, Cambridge, MA 02138, USA}
\altaffiltext{15}{The Adler Planetarium, Chicago, IL 60605, USA}
\altaffiltext{16}{School of Physics \& Astronomy, University of Manchester, Oxford Road, Manchester, M13 9PL, UK}
\altaffiltext{17}{c/o Australian Broadcasting Corporation}
\altaffiltext{18}{Department of Astrophysics, California Institute of Technology, Pasadena, CA 91125, USA}
\altaffiltext{19}{The University of Tokyo, 7-3-1 Bunkyo-ku, Tokyo 113-0033, Japan}
\altaffiltext{20}{Institute for Astronomy, University of Hawai`i at M\={a}noa, Honolulu, HI 96822, USA}
\altaffiltext{21}{Arizona State University, Tempe, AZ 85281, USA}
\altaffiltext{22}{Citizen Scientists, c/o Zooniverse, Department of Physics, University of Oxford, Denys Wilkinson Building, Keble Road, Oxford, OX1 3RH, UK}
\altaffiltext{23}{Danmarks Tekniske Universitet, Anker Engelundsvej 1, Building 101A, 2800 Kgs. Lyngby, Denmark}

\begin{abstract}
\ky\ is a moderately bright ($V=12.2$, $K=10.3$) main sequence K-star observed in Campaign 12 of the NASA \emph{K2} mission. It hosts five small (1.6--3.3R$_{\oplus}$) transiting planets in a compact architecture. The periods of the five planets are 2.35~d, 3.56~d, 5.40~d, 8.26~d, and 12.76~d, forming an unbroken chain of near 3:2 resonances. Although we do not detect the predicted 2--5 minute transit timing variations with the \emph{K2} timing precision, they may be observable by higher cadence observations with, for example, \emph{Spitzer} or CHEOPS. The planets are amenable to mass measurement by precision radial velocity measurements, and therefore \ky\ could represent a new benchmark systems for comparing radial velocity and TTV masses. K2-138 is the first exoplanet discovery by citizen scientists participating in the Exoplanet Explorers project on the Zooniverse platform.
\end{abstract}


\section{Introduction}

The NASA \emph{K2} mission \citep{Howell2014} is in its third year of surveying the ecliptic plane. The mission uses the repurposed \emph{Kepler Space Telescope} to tile the ecliptic, and consists of successive $\sim$80-day observations of 12$\times$12 degree regions of sky known as campaigns. Each campaign yields high-precision, high-cadence calibrated pixel files and time-series photometry on anywhere from 13,000 to 28,000 targets, which are released to the public within three months of the end of the observing campaign. This deluge of data is immediately inspected by professional exoplanet hunters, producing rapid announcements of interesting new planetary systems, e.g. the recent examples of HD~106315 \citep{Crossfield2017,Rodriguez2017} and HD~3167 \citep{Vanderburg2016}.

There are many features in time-series data which can be matched to potential transit signals by signal-processing algorithms. These features can be either astrophysical in origin (e.g. pulsating variable stars, eclipsing binaries, flaring stars, cosmic-ray pixel strikes), or instrumental (e.g. apparent variations in the brightness in a photometric aperture caused by spacecraft pointing drift, or by focus drifts in response to the changing thermal environment). While these artifacts can confuse automated procedures, the human brain is optimised for pattern matching, and is remarkably good at discriminating these artifacts from a train of planet transits. This ability is exploited in information security technology for instance, with the CAPTCHA algorithm \citep{vonAhn2003} being a well-known example. In light curve analysis, humans can readily recognise the differences in the underlying phenomena causing the putative signals and identify the transit signals.

This ability, along with the strong interest held by the public in being involved in the process of scientific discovery, has led to the ongoing success of the Planet Hunters\footnote{\url{www.planethunters.org}} project \citep[e.g.][]{Fischer2012}. Hosted by the Zooniverse platform \citep{Lintott2008}, the project displays to users the publicly available \emph{Kepler} and \emph{K2} time series photometry, and asks them to identify transit-like dips. Inspired by their success, we started the Exoplanet Explorers\footnote{\url{www.exoplanetexplorers.org}} project in April 2017. In this project we run a signal detection algorithm to identify potential transit signals in the \emph{K2} time series photometry, and ask the users to sift through the resulting candidates to identify those most closely resembling planetary transits. Here we present K2-138, the first \emph{K2} planetary system discovered by citizen scientists. In Section \ref{sec:obs}, we describe the \emph{K2} data set. We describe the Exoplanet Explorers project and the identification of K2-138 in Section \ref{sec:results}, and the derivation of the stellar parameters in Section \ref{sec:stellar}. In Section \ref{sec:parameters} we describe the analysis of the planet parameters. Finally, in Section \ref{sec:disc} we place the K2-138 system in context of other high-multiplicity systems and discuss prospects for future characterisation.

\section{Observations and Photometric Reduction}
\label{sec:obs}


\begin{figure}
\plotone{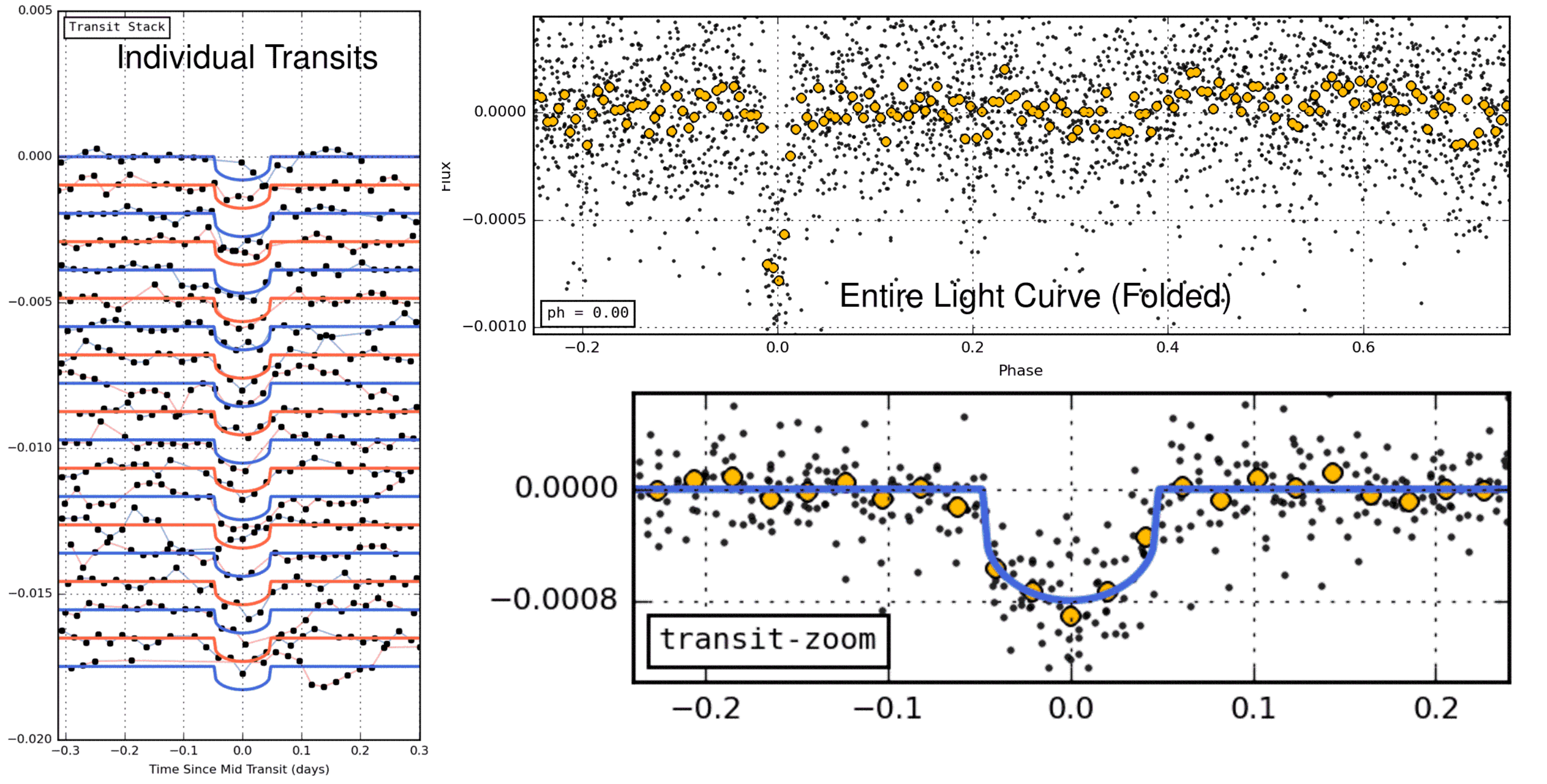}
\plotone{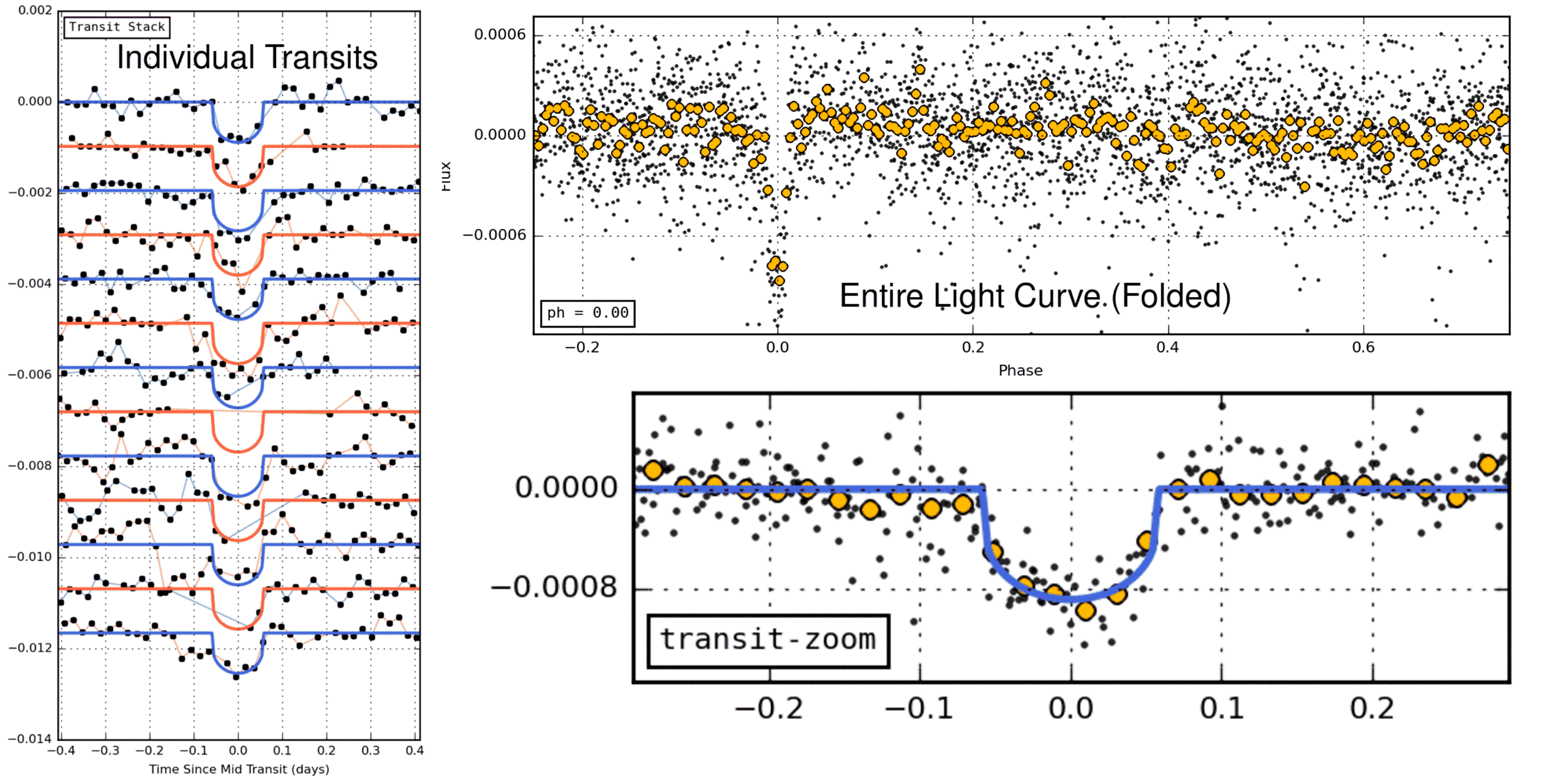}
\plotone{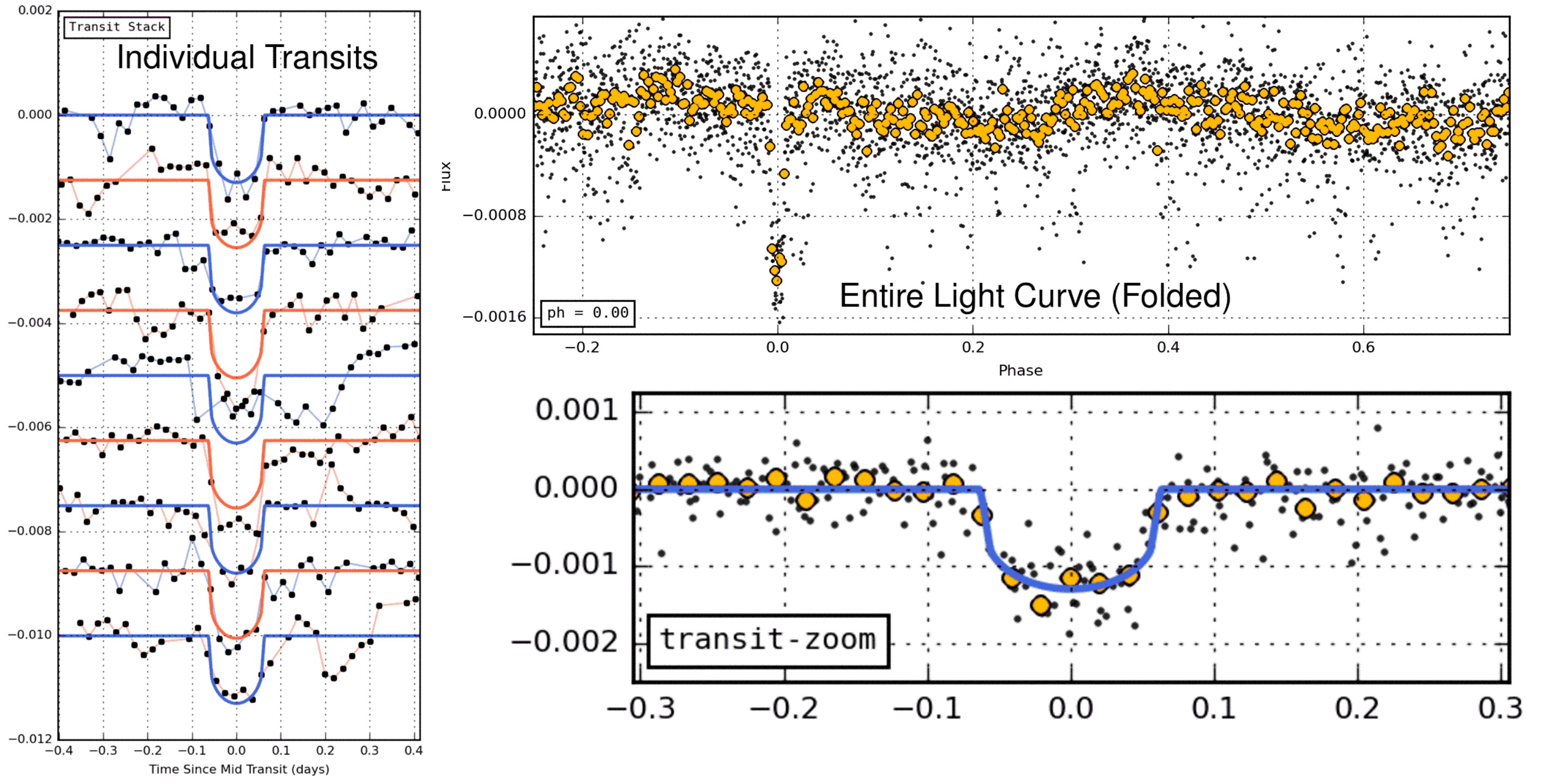}
\plotone{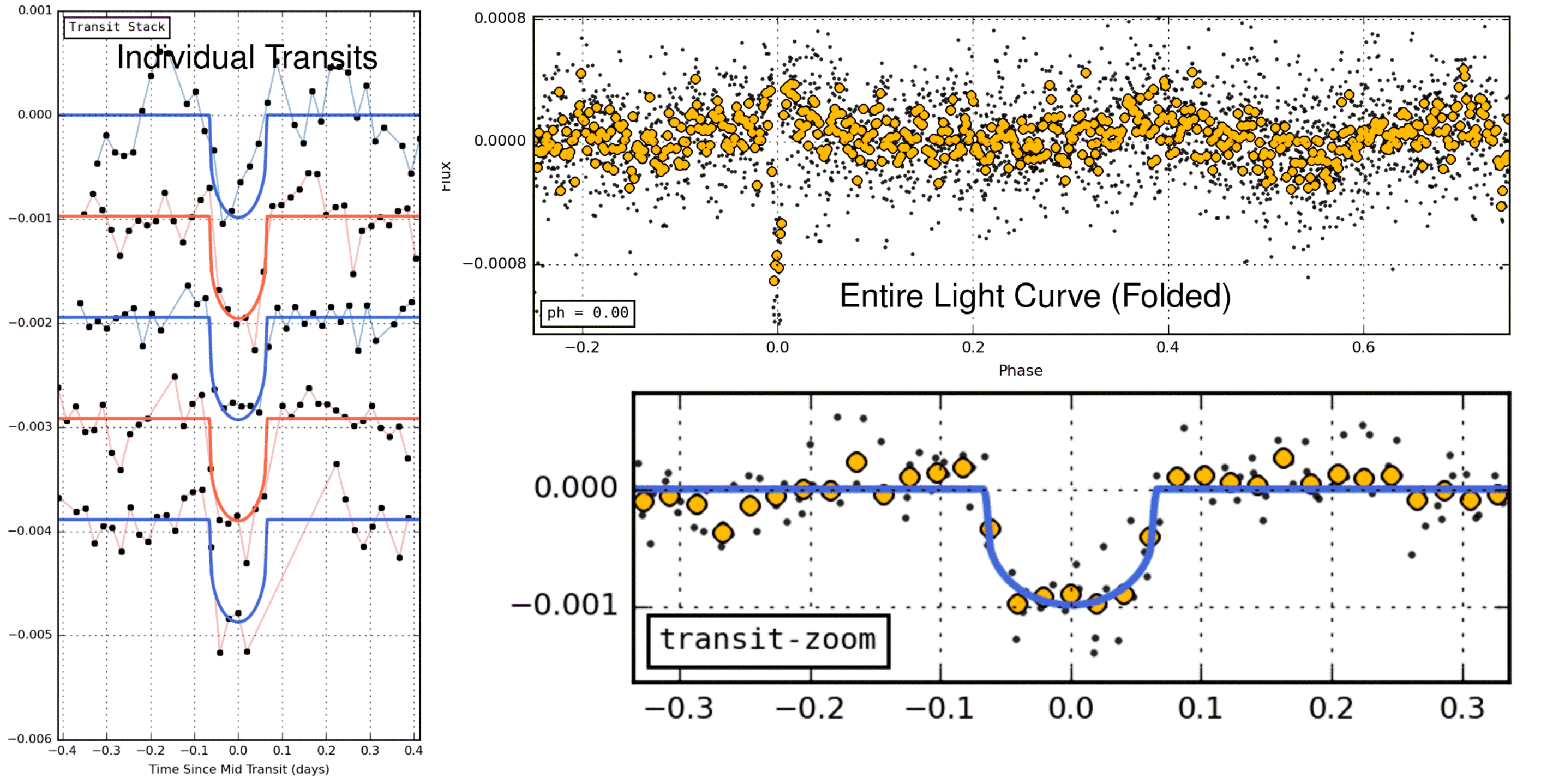}
\caption{The set of diagnostic plots presented on the Exoplanet Explorers project, for the target K2-138. From top to bottom the four plots are for the putative signals for K2-138 c, d, e, and f. In each case, the left panel shows the individual transit events, with an arbitrary vertical offset and alternating color for visual clarity. The top right panel shows the entire light curve folded at the period of the putative transit signal. The black points are the original \emph{K2} data, and the yellow circles are binned data. The bottom right panel shows the same phase-folded light curve, zoomed in on the transit event itself. An initial fitted planet model is overlaid in blue. \label{fig:zooniverse}}
\end{figure}

\emph{K2} data are downlinked from the spacecraft, processed into calibrated pixel files and photometric time series, and released to the public via the Mikulski Archive for Space Telescopes (MAST)\footnote{\url{https://archive.stsci.edu/kepler/}}. Unlike the original \emph{Kepler} mission, the target list is entirely guest observer driven: all observed targets are proposed to the project by the community. Campaign 12 (C12), which was observed for 79 days from 2016 December 15 to 2017 March 4, contained the target star TRAPPIST-1 \citep{Gillon2016,Gillon2017,Luger2017,Wang2017}. In order to facilitate rapid analysis of the \emph{K2} observations of TRAPPIST-1, the data were released to the public immediately after downlink from the spacecraft on 2017 March 9 as raw pixel files.

We downloaded the raw pixel files from MAST for the stellar targets that were proposed by K2's large exoplanet search programs, and calibrated these data using the \texttt{kadenza} software \citep{Barentsen2017}, generating calibrated pixel files.


We then used the publicly available \texttt{k2phot} photometry code\footnote{\url{https://github.com/petigura/k2phot}}, which generates aperture photometry and performs corrections for the spacecraft pointing jitter using Gaussian Processes \citep{Rasmussen2005}, to generate light curves suitable for searching for periodic transit signals. Using the publicly available TERRA algorithm\footnote{\url{https://github.com/petigura/terra}} \citep{Petigura2013a,Petigura2013b}, we generated both a list of potential transiting planet signals from the detrended light curves, and a set of accompanying diagnostic plots. TERRA identified a total of 4,900 candidate transit signals in the C12 stellar data.

\section{Transit Identification}
\label{sec:results}

For each signal from C12 and the earlier campaigns already processed, we uploaded a subset of the standard TERRA diagnostic plots to the Exoplanet Explorers project. The plots included a phase-folded light curve and a stack of the individual transit events; users examined these plots and selected whether the putative signal looked like a true transiting planet candidate. Figure \ref{fig:zooniverse} shows an example set of diagnostic plots. 

On 2017 April 4, the Exoplanet Explorers project was featured on the Stargazing Live ABC broadcast in Australia. Between April 4, 2017 01:00 UTC and April 6, 2017 19:48 UTC, the live project received 2,100,643 classifications from 7,270 registered Zooniverse classifiers and 7,677 not-logged-in IP addresses. Of these, 130,365 classifications from 4,325 registered classifiers and 2,012 not-logged-in IP addresses were for candidate signals in the C12 data. C12 candidates received a median of 26 classifications each; the C12 candidate with the lowest classification count received 14 classifications, and the most-classified C12 candidate received 43 classifications. The classifications were aggregated for each candidate to provide the fraction of classifiers who indicated they saw a transiting planet signal. 

\begin{figure}
\epsscale{1.15}
\plotone{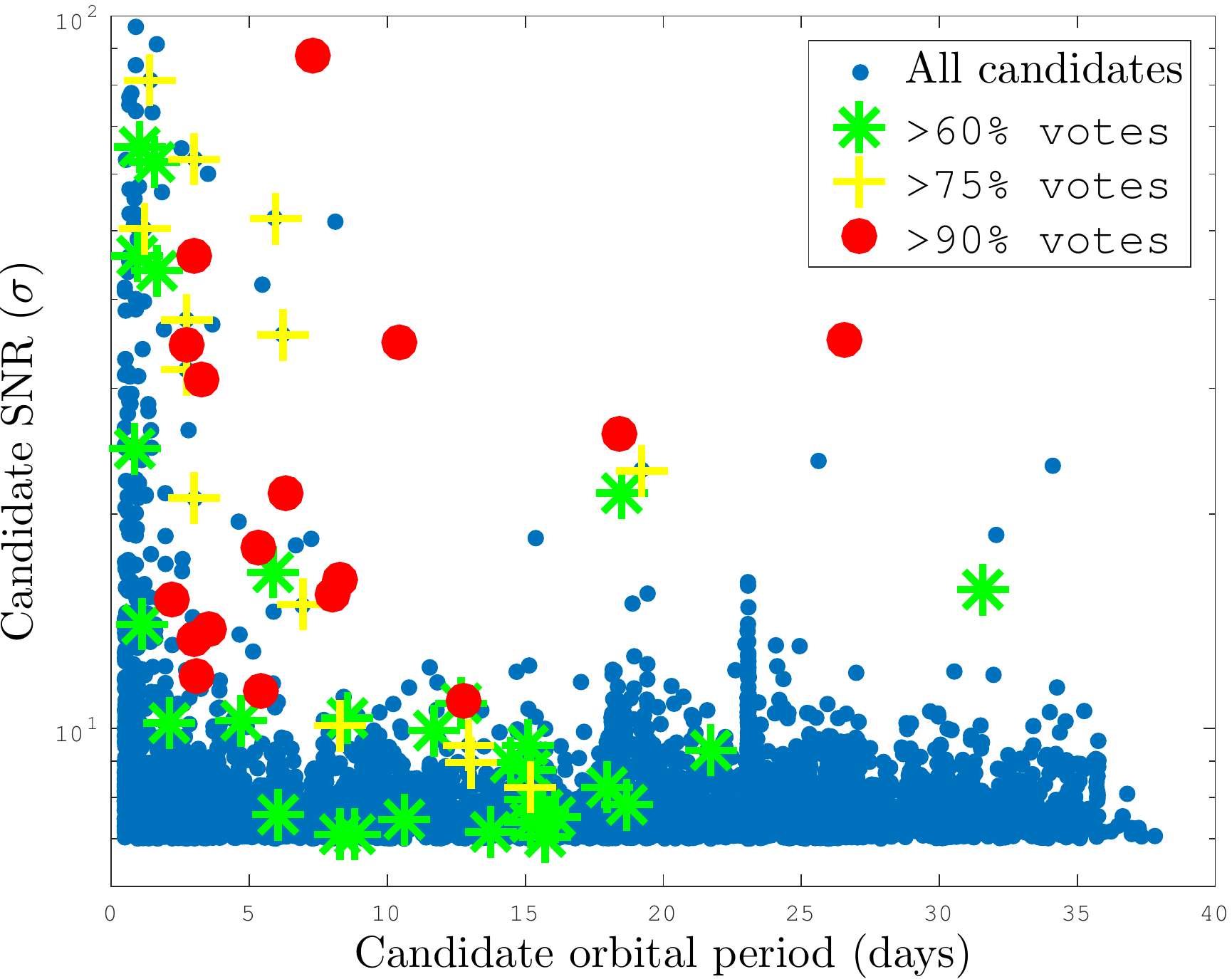}
\caption{The distribution of user votes on the C12 candidate transiting planet signals. All signals received 14 or more votes, and the percentage of `yes' votes is shown: the small blue dots represent signals for which fewer than 60\% of users voted `yes'. The green stars, yellow $+$ symbols and red circles show signals for which greater than 60\%, 75\% and 90\% of users voted `yes'. \label{fig:dispositions}}
\end{figure}

Of the 4,900 potential transiting signals identified in the C12 data, 72 were voted by more than 60\% of users as looking like transiting planet candidates. The dispositions of the full set of C12 signals are shown in Figure \ref{fig:dispositions}; the signals with the highest confidence are unsurprisingly at shorter periods (with a higher number of individual events contributing to the signal for the users to assess) and higher signal-to-noise values. The highly voted signals were inspected visually and \ky\ (EPIC~245950175) was rapidly identified as a promising multi-planet system. The initial automated search of the light curve produced four distinct transiting signatures, each with a high ($>$90\%) fraction of votes from the participants; the four diagnostic plots that were voted on are shown in Figure \ref{fig:zooniverse}. EPIC~245950175 was proposed for observations by four teams, in Guest Observing Programs 12049, 12071, 12083, and 12122 (PIs Quintana, Charbonneau, Jensen, and Howard). The full, unfolded light curve of \ky\  is shown in Figure \ref{fig:lc}. After the system was flagged by the citizen scientists, additional examination of the light curve revealed the signature of a fifth transiting signal, interior to the four signals identified by the TERRA algorithm and on the same 3:2 resonant chain. Characterisation of the five detected planet signals is detailed in Section \ref{sec:parameters}.

\begin{figure}
\epsscale{1.25}
\plotone{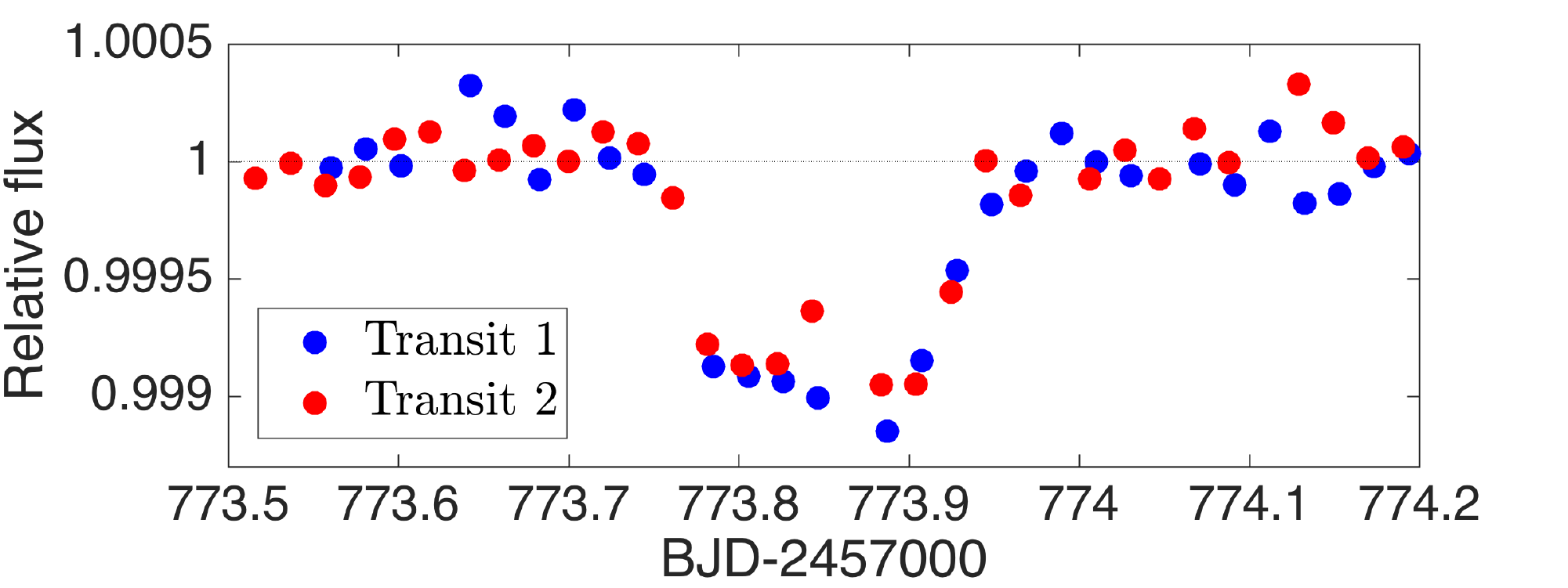}
\caption{The two transits of the putative 42-day period planet candidate that fall in the \emph{K2} C12 data. The time of the second transit has been offset by 41.97 days to demonstrate the consistency in depth and duration of the two events, which correspond to a $\sim$2.8R$_{\oplus}$ planet. \label{fig:candidate}}
\end{figure}

\begin{figure*}
\plotone{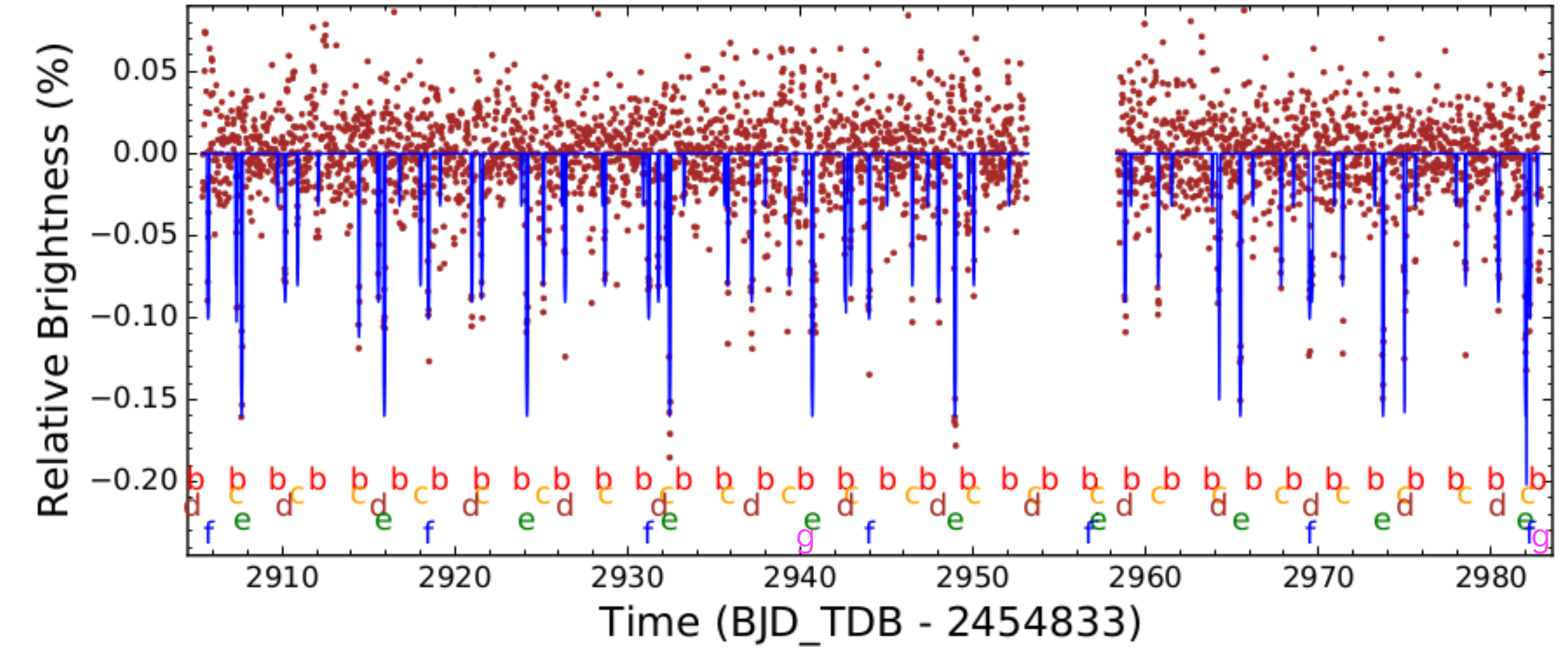}
\caption{The time series of the K2 data, with the five-planet transit model shown in blue. The planets of the individual transits are marked with the appropriate letter; the times of the two transits of the putative planet candidate discussed in Section \ref{sec:results} are shown as `g'. The 5-day gap two-thirds of the way through the campaign was the result of a spacecraft safe-mode event. \label{fig:lc}}
\end{figure*}

In addition, two individual transit events were identified, shown in Figure \ref{fig:candidate}, separated by 41.97 days. The transits have consistent depths and durations, and if confirmed, would correspond to an additional $\sim$2.8R$_{\oplus}$ sub-Neptune planet in the \ky\ system, bringing the total to six planets. Additional observations are required to secure a third epoch and confirm that the two transits seen in the \emph{K2} data arise from a single planet, and are not individual transits of two similarly-sized, longer-period planets.

\section{Stellar Characterisation}
\label{sec:stellar}

On 2017 June 1 we obtained a spectrum of \ky\ using Keck/HIRES, without the iodine cell as is typical of the precision radial velocity observations. We derive stellar parameters using SpecMatch \citep{Petigura2015}, given in Table \ref{tab:stellar}. Following the procedure in \citet{Crossfield2016}, we estimate the stellar radius and mass using the publicly available {\texttt isochrones} Python package \citep{Morton2015} and the Dartmouth stellar evolution models \citep{Dotter2008}. The California Kepler Survey \citep[CKS;][]{Petigura2017} finds mass and radius uncertainty floors for similar spectral types of 6\% and 9\% respectively, motivated by comparisons between stellar radii derived using \texttt{isochrones} and spectroscopic parameters (as is done here for \ky) and asteroseismic radii \citep{Johnson2017}. We therefore adopt these uncertainties on the mass and radius of \ky\ to incorporate the isochrone model uncertainties. Additional estimates of the parameters of \ky\ are available in the Ecliptic Plane Input Catalog \citep[EPIC,][]{Huber2016} on the MAST and are consistent with those from the RAVE spectrum and \texttt{isochrones}. The HIRES/\texttt{isochrones} parameters and the EPIC parameters are consistent with a solar-metallicity, main-sequence, early-K type star at a distance of $\sim$180 pc when comparing to the color-temperature relations of \citet{Pecaut2013}. We adopt a spectral type of K1V~$\pm$~1. We measure a log~$R\prime_{HK}$ value of $-4.63$, indicating a modestly magnetically active star, which may present a challenge for precision radial velocity measurements of the system.\\

\begin{figure}
\epsscale{1.0}
\plotone{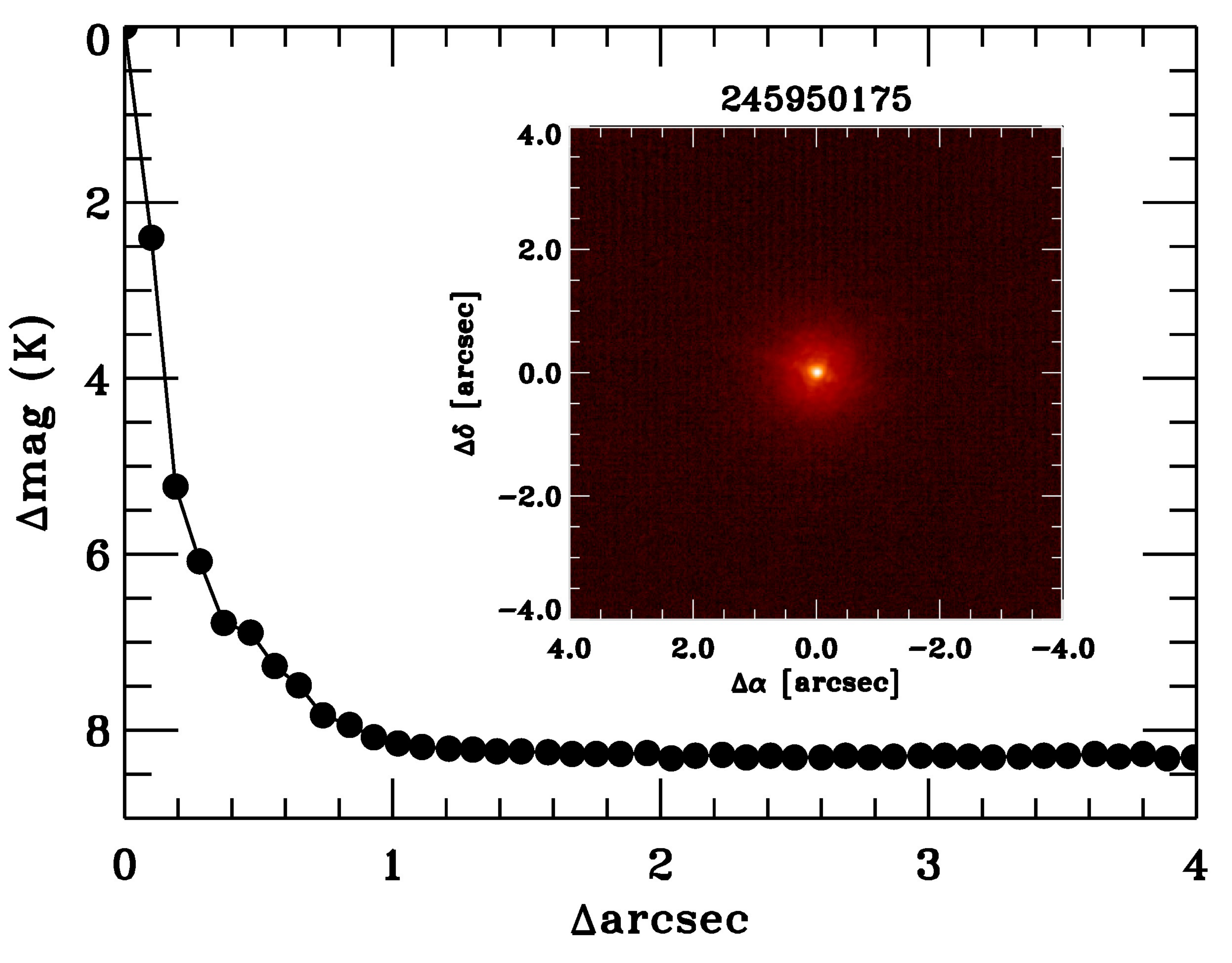}
\caption{\emph{Inset:} The Gemini/NIRI AO image of \ky. We detect no additional sources of flux. \emph{Plot:} The 5-$\sigma$ contrast limits for additional companions, in $\Delta$magnitude, are plotted against angular separation in arcseconds; the black points represent one step in the FWHM resolution of the images.\label{fig:ao}}
\end{figure}

On 2017 May 31 we obtained a high-resolution image of \ky\ in $K$-band using the Altair AO system on the NIRI camera at Gemini Observatory \citep{Hodapp2003} under program GN-2105B-LP-5 (PI Crossfield). We observed at five dither positions, and used the dithered images to remove sky background and dark current; we then aligned, flat-fielded and stacked the individual images. The inset in Figure \ref{fig:ao} shows the final stacked image, and the plot shows the detection limits of the final image. The limits were determined by injecting simulated sources into the final image, with separation from \ky\ determined by integer multiples of the FWHM, as in \citet{Furlan2017}. We see no other source of contaminating flux in the AO image within 4$^{\prime\prime}$, the size of one \emph{K2} pixel. In addition to the AO data, we examine the HIRES spectrum for evidence of additional stellar lines, following the procedure of \citet{Crossfield2016}. This method is
sensitive to secondary stars that lie within 0.4$^{\prime\prime}$ of the primary star (one half of the slit width) and that are up to 5 magnitudes fainter than  the primary star in the $V$- and $R$-bands \citep{Kolbl2015}, complementing the sensitivity limits of the NIRI observations. We are able to rule out companions with T$_{\rm{eff}}$ = 3400--6100 K and $\Delta$(RV)$>$10 kms$^{-1}$. We further discuss the possibility of the observed periodic signals originating from a faint star $\sim$14$^{\prime\prime}$ away in Section \ref{sec:validation}, but for the following analysis we assume the putative planet signals arise from \ky.\\

\begin{table}
\footnotesize
\begin{center}
\caption{\ky\  stellar parameters \label{tab:stellar}}
\begin{tabular}{ll}
\hline
EPIC ID & 245950175\\
2MASS ID & J23154776-1050590 \\
RA (J2000.0) & 23:15:47.77 \\
Dec (J2000.0) & -10:50:58.91 \\
$V$ (mag) & 12.21 \\
$K$ (mag) & 10.305 \\
Spectral type & K1V $\pm$ 1\\
T$_{\rm{eff}}$ (K) & 5378 $\pm$ 60\\
log $g$ (cgs) & 4.59 $\pm$ 0.07 \\
\lbrack Fe/H\rbrack & 0.16 $\pm$ 0.04 \\
R$_\star$ (R$_{\odot}$) & 0.86 $\pm$ 0.08\\
M$_\star$ (M$_{\odot}$) & 0.93 $\pm$ 0.06\\
Distance (pc)\footnote{EPIC classification, see \citet{Huber2016} and \texttt{https://github.com/danxhuber/galclassify} } & 183 $\pm$ 17 \\
$v$ sin $i$ (km/s) & 2.7 $\pm$ 1.5 \\
\tableline
\end{tabular}
\end{center}
\end{table}

\section{Planet Parameters}
\label{sec:parameters}

\begin{figure}
\epsscale{1.0}
\plotone{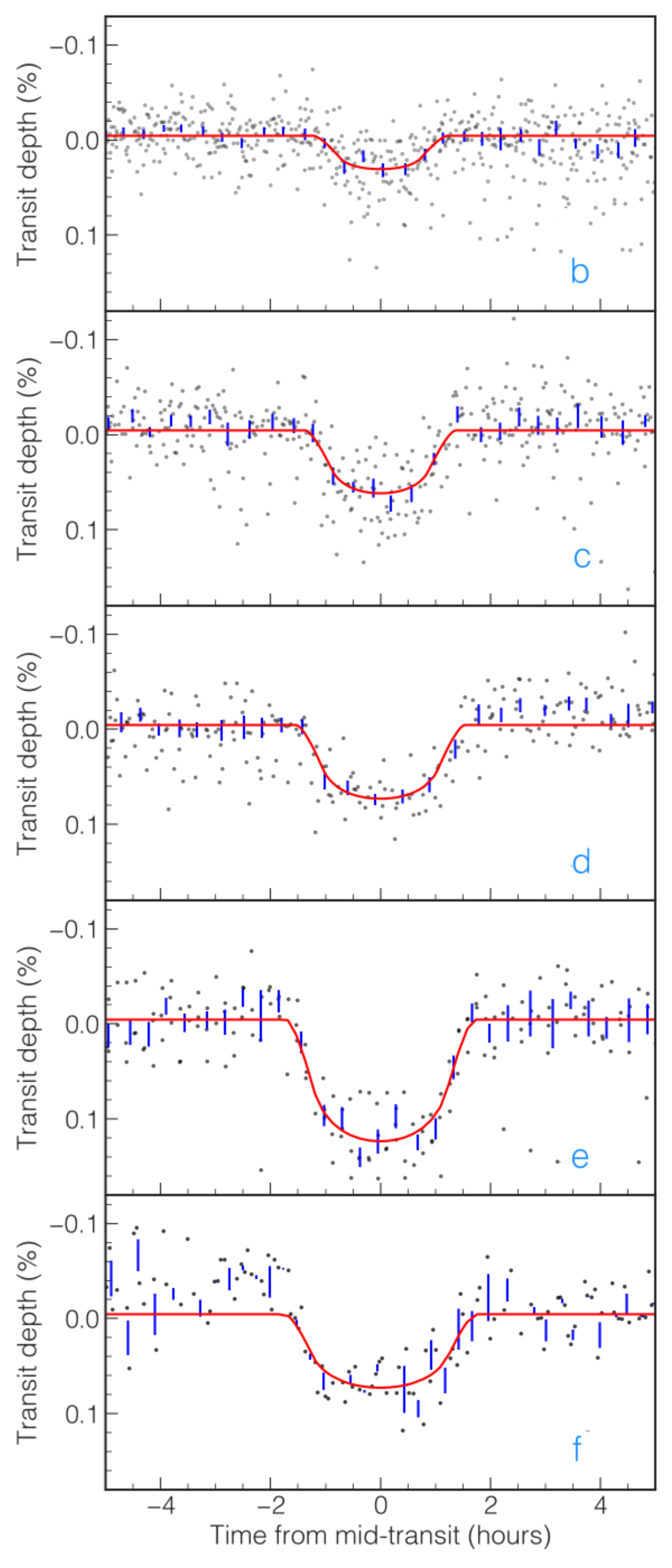}
\caption{The folded transits of \ky\ b, c, d, e, and f overplotted with the best fitting transit model in red. Binned data points are shown in blue. The planets range in size from 1.57 to 3.29 R$_{\oplus}$. \label{fig:transitmodel}}
\end{figure}

We analyzed the five transit signals independently in the \ky\ light curve, using the same modeling, fitting, and MCMC procedures as described in \citet{Crossfield2016}. In summary, we fit the following model parameters: mid-transit time (T$_0$); the candidate’s orbital period and inclination (P and i); the scaled semimajor axis (R$_p$/a); the fractional candidate size (R$_*$/a); the orbital eccentricity and longitude of periastron (e and $\omega$), the fractional level of dilution ($\delta$) from any other sources in the aperture; a single multiplicative offset for the absolute flux level; and quadratic limb-darkening coefficients (u1 and u2). We explore the posterior distribution using the \texttt{emcee} software \citep{Foreman2013}. We find that the signals correspond to five sub-Neptune-sized planets ranging from 1.6--3.3R$_{\oplus}$; the best fitting transit models are shown in Figure~\ref{fig:transitmodel}. As a self-consistency check, we note that the stellar density values derived from the independent transit fits are consistent across all five planets, and also consistent with the direct calculation from the stellar mass and radius.\\

All five planets have periods under 13 days, making \ky\ an example of a tightly-packed system of small planets. One particularly interesting aspect to the \ky\  architecture, discussed further in Section \ref{sec:disc}, is that each successive pair of planets is just outside the first-order 3:2 resonance. 

\subsection{Validation}
\label{sec:validation}
\citet{Lissauer2012} analysed the distribution of \emph{Kepler} planet candidates and showed that systems with multiple candidate signals were substantially more likely to be true planetary systems than false positives. This provides a `multiplicity boost' to the statistical validation of candidates in multi-planet systems. Here, we validated each candidate individually using the publicly available \texttt{vespa} code\footnote{\url{https://github.com/timothydmorton/VESPA}}, which computes the likelihood of various astrophysical false positive scenarios. We use as input a photometric exclusion radius of  13$^{\prime\prime}$, the $K$-band and \emph{Kepler} magnitudes, and the HIRES stellar parameters. The results are false positive probabilities of 0.20\%, 0.11\%, 0.76\%, 0.027\%, and 0.24\% for planets b, c, d, e, and f respectively. Since we have multiple candidates orbiting a single star, the `multiplicity boost' (and an additional `near-resonance boost') further suppresses these FPPs \citep{Lissauer2012,Sinukoff2016}. Applying the \emph{K2} multiplicity boost derived by \citet{Sinukoff2016}, we find final FPPs of 8.3$\times 10^{-5}$, 4.6$\times 10^{-5}$, 3.17$\times 10^{-4}$, 1.1$\times 10^{-5}$, and 1.0$\times 10^{-4}$. 

\begin{figure}
\plotone{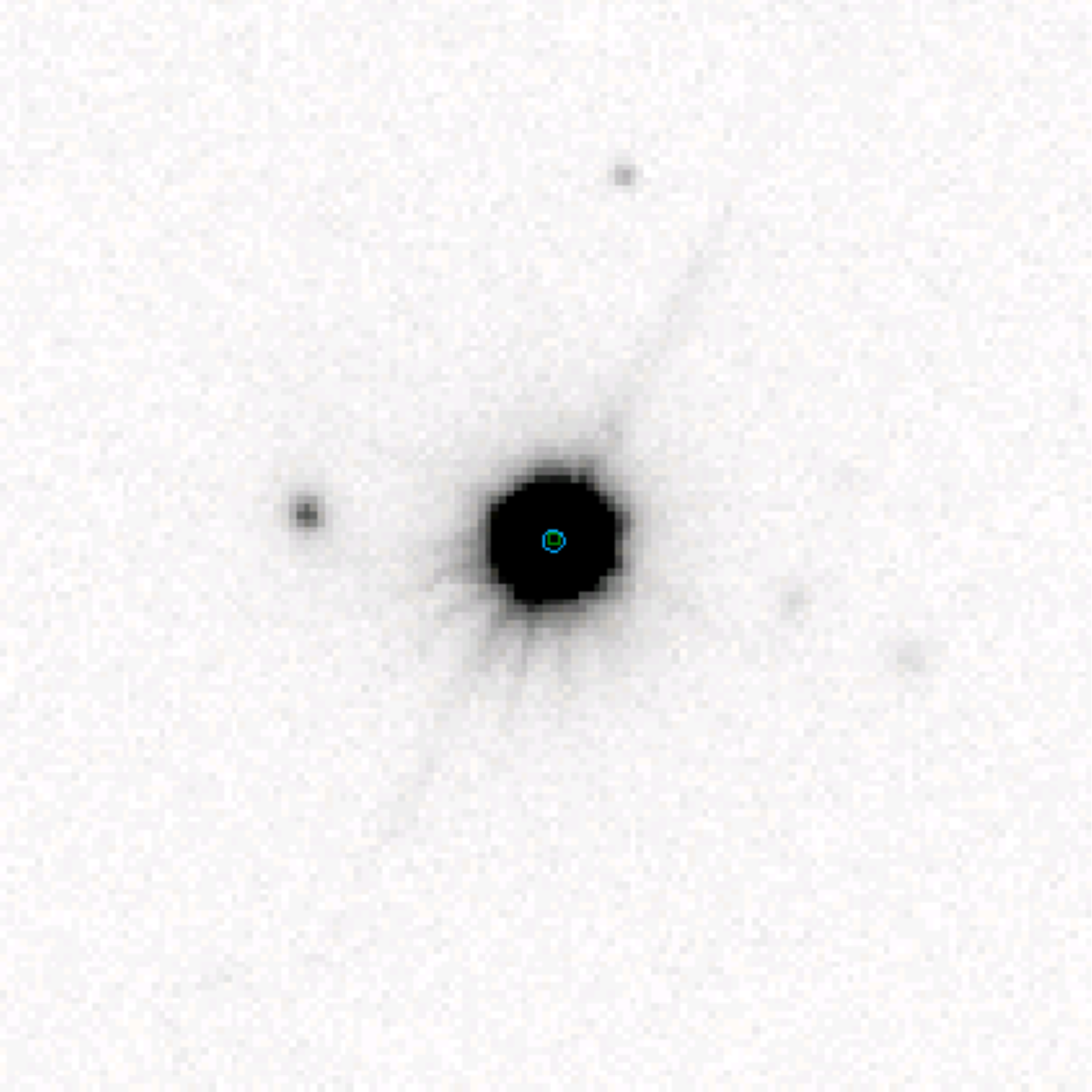}
\caption{A 60$\times$60 arcsecond image from the SDSS DR7 $r$-band. The companion to the west is $\sim$14 arcseconds away, and is 5.6 magnitudes fainter than EPIC 245950175 in $R$. \label{fig:finderchart}}
\end{figure}

Recently, \citet{Cabrera2017} showed that stars within the \emph{Kepler} photometry aperture but outside the small area surveyed by high-resolution imaging were responsible for several falsely validated planets. Here we examine the possibility that the five periodic signals do not arise from the brightest star in the \emph{K2} aperture. Given that the signals form an unbroken chain of near first-order resonances, we consider the possibility that some number of the signals arise on one star and the remainder on a star coincident with the line of sight to be unlikely, and consider the five signals as a related set. The brightest nearby star is 2MASS~J23154868-1050583, which is $\sim$14$^{\prime\prime}$ from EPIC~245950175, and 5.6 magnitudes fainter in $R$-band. This star is shown to the west of EPIC~245950175 in Figure \ref{fig:finderchart}. Following from Eq. (5) of \citet{Ciardi2015}, we find that the putative planets would be 13.2 times larger if they orbited the fainter target, increasing to 1.9--4.0 R$_{\rm{J}}$. These would be as large or larger than the largest planet known to date with a radius measured by the transit method, WASP-79b with a radius of 2.09$\pm$0.14 R$_{\rm{J}}$ \citep{Smalley2012}. 
Therefore we conclude that the five putative planets are extremely unlikely to orbit 2MASS~J23154868-1050583. 

\section{Discussion}
\label{sec:disc}

One of the interesting discoveries from the NASA \emph{Kepler} mission is the prevalence of compact, highly co-planar, and often dynamically packed systems of small ($<4R_\oplus$) planets \citep{Latham2011,Lissauer2011,Fabrycky2014,Howard2012,Winn2015}. This has continued in the \emph{K2} mission, including the discoveries of the K2-3 \citep{Crossfield2015}, K2-37 \citep{Sinukoff2016}, and K2-72 \citep{Crossfield2016} systems. Multi-planet systems are crucial laboratories for testing planetary formation, migration and evolution theories. A further interesting subset of these systems are those demonstrating resonances, or chains of resonances. The five validated planets of \ky\ lie close to a first-order resonant chain. We find period ratios of 1.513, 1.518, 1.528 and 1.544 for the b-c, c-d, d-e, and e-f pairs respectively, just outside the 3:2 resonance. \citet{Fabrycky2014}, examining the large population of multi-transiting planet systems in the \emph{Kepler} data, showed that pair-wise period ratios pile-up just outside of the first- and second-order resonances. They offer several possible explanations for this, including gravitational scattering slightly out of resonance by the additional bodies in the system, or tidal dissipation preferentially acting to drag the inner planets inward from the resonance. \citet{Lithwick2012} and \citet{Batygin2013} investigate the suggestion of tidal dissipation as a mechanism for keeping individual pairs of planets just outward of the resonance; they note that in systems with more than two planets, where the planets can inhabit multiple resonances, the planets can remain close to resonance despite tidal dissipation. Recently, \citet{Ramos2017} analytically derived the expected offset from a first-order resonance for a pair of planets due to Type I migration. Their Fig. 3 shows that for periods shorter than $\sim$10 days, the resonance period ratio is 1.505--1.525, depending on the mass of the inner planet and the mass ratio of the two planets, with higher period ratios expected as the mass ratio approaches unity. Therefore it is possible that the \ky\ b-c and c-d pairs may be captured in the 3:2 resonance, but unlikely that the d-e or e-f pairs are in resonance.\\

\begin{figure}
\epsscale{1.15}
\plotone{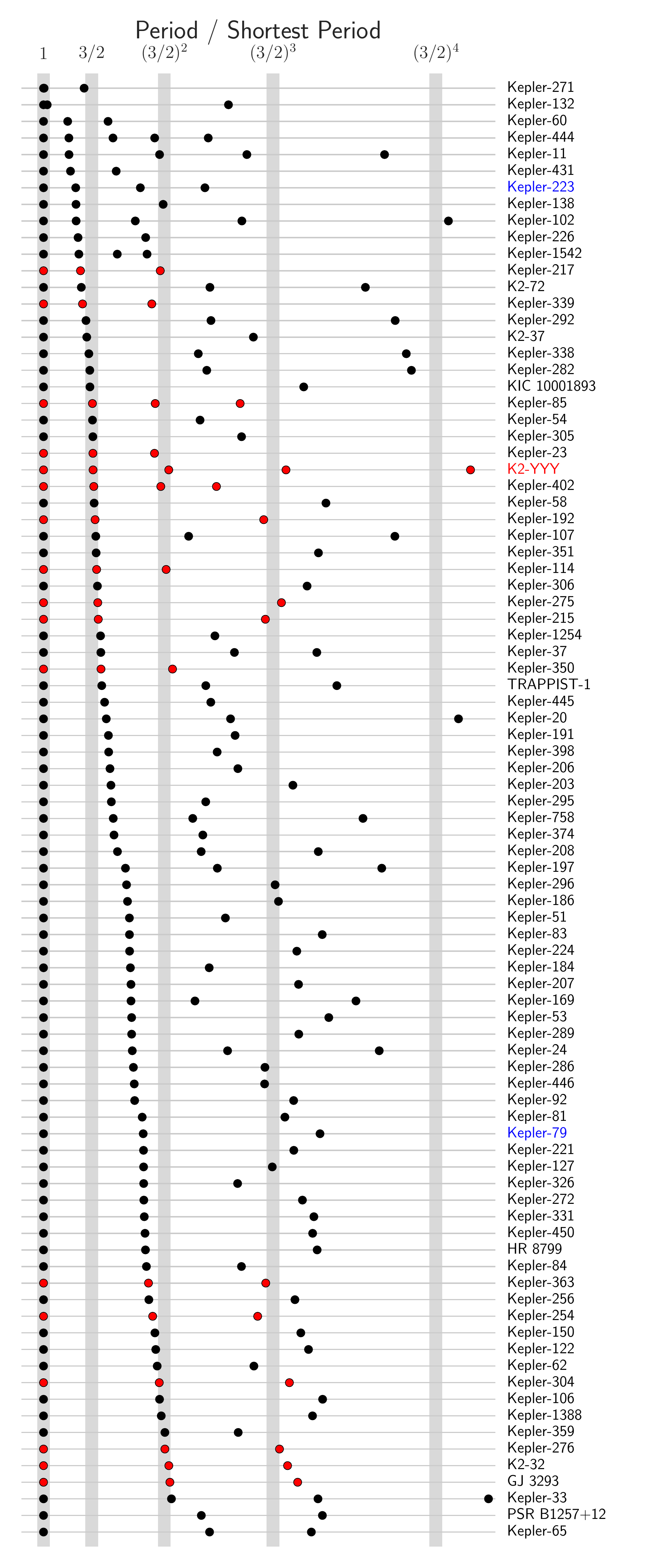}
\caption{The distribution of distances from 3:2 period ratios in confirmed multi-planet systems that have three or more planets in a compact geometry (defined as having three planets with Period/Shortest Period $<$ 4). Planetary systems with multiple near-3:2 resonances are highlighted in red. K2-138 is the only system near an unbroken chain of four near-3:2 resonances. Kepler-79 and Kepler-223 (shown in blue) both have four planets in or near a chain of resonances. The vertical lines indicate the positions of successive 3:2 period ratios. \label{fig:multisystems}}
\end{figure}

It is illustrative to compare \ky\ to the other known systems with multiple planets, and to examine whether 3:2 period ratios are common. Figure \ref{fig:multisystems} shows, for the confirmed multi-planet systems, the distance of each period ratio in the system from a 3:2 period ratio. K2-138 is the only system with an unbroken chain of four period ratios near 3:2. There are seven systems with planets in consecutive 3:2 pairs: Kepler-23 \citep{Ford2012}; Kepler-85 and Kepler-114 \citep{Xie2013,Rowe2014}; Kepler-217 \citep{Rowe2014,Morton2016}; Kepler-339 and Kepler-402 \citep{Rowe2014}; and Kepler-350 \citep{Rowe2014,Xie2014}. There are an additional nine systems with a `broken' chain of 3:2 pairs, i.e. a configuration like K2-138 but where one planet is missing, or perhaps undiscovered: GJ 3293 \citep{Astudillo2015,Astudillo2017}; K2-32 \citep{Dai2016}; Kepler-192 and Kepler-304 \citep{Rowe2014,Morton2016}; Kepler-215, Kepler-254, Kepler-275 and Kepler-363 \citep{Rowe2014}; and Kepler-276 \citep{Rowe2014,Xie2014}. These systems are highlighted in red. 
A small number of systems contain four planets in different configurations of first order resonances, also highlighted in Figure \ref{fig:multisystems}. Kepler-223 is comparable to \ky\ : a compact system of four sub-Neptune-sized planets with periods shorter than 20 days, in a 3:4:6:8 resonant chain \citep{Rowe2014,Mills2016}. In the case of Kepler-223, the period ratios are much closer to resonance than for \ky, with ratios of 1.3333, 1.5021, and 1.3338 for the b-c, c-d, and d-e pairs respectively. Kepler-223 demonstrates significant transit timing variations, allowing for robust mass constraints to be placed. Kepler-79 \citep{Jontof-Hutter2014} is a scaled-up version of \ky\  and Kepler-223, with four sub-Saturn-sized planets in a 1:2:4:6 resonant chain with periods from 13--81 days. Finally, the benchmark TRAPPIST-1 system hosts seven planets in a resonant chain, with successive period ratios of 8:5, 5:3, 3:2, 3:2, 4:3, and 3:2 \citep{Gillon2017,Luger2017}. Like TRAPPIST-1, \ky\ may represent a pristine chain of resonances indicative of slow, inward disk migration. 

Another notable feature of the TRAPPIST-1 system is that the seven known planets form a complex chain of linked three-body Laplace resonances \citep{Luger2017}. Similarly, Kepler-80 (KOI-500) is a five-planet system where the four outer planets form a tightly linked pair of three-body resonances \citep{Lissauer2011,MacDonald2016}; Kepler-223, described above, also contains a pair of three-body resonances. One other system, Kepler-60, appears to be in either a true three-body Laplace resonance or a chain of two-planet mean motion resonances \citep{Gozd2016}. A three-body resonance satisfies the condition that $(p/P_1) - [(p+q)/P_2] + (q/P_3) \approx 0$, where $p$ and $q$ are integers and $P_i$ the period of the $i$th planet. For \ky\, we find that the three consecutive sets of three planets (bcd, cde, and def) all satisfy this condition with $(p,q)=(2,3)$, resulting in values of 4.2$\pm$1.7$\times$10$^{-4}$ days$^{-1}$, $-1.6\pm$0.9$\times$10$^{-4}$ days$^{-1}$, and $-2.4\pm$4.9$\times$10$^{-4}$ days$^{-1}$ respectively, all close to zero. In simulating the  Kepler-80 system, \citet{MacDonald2016} find that their migration simulations can naturally describe the final system architecture, with dissipative forces pushing the interlocked planets out of two-body resonances and into three-body resonances; the \ky\ system may have undergone something similar. \ky\ joins a relatively modest population of known systems with four or more planets in or close to a resonant chain, and a very small population of systems with interlocking chains of three-body resonances, making it an ideal target to study for transit timing variations.


\begin{figure}
\epsscale{1.25}
\plotone{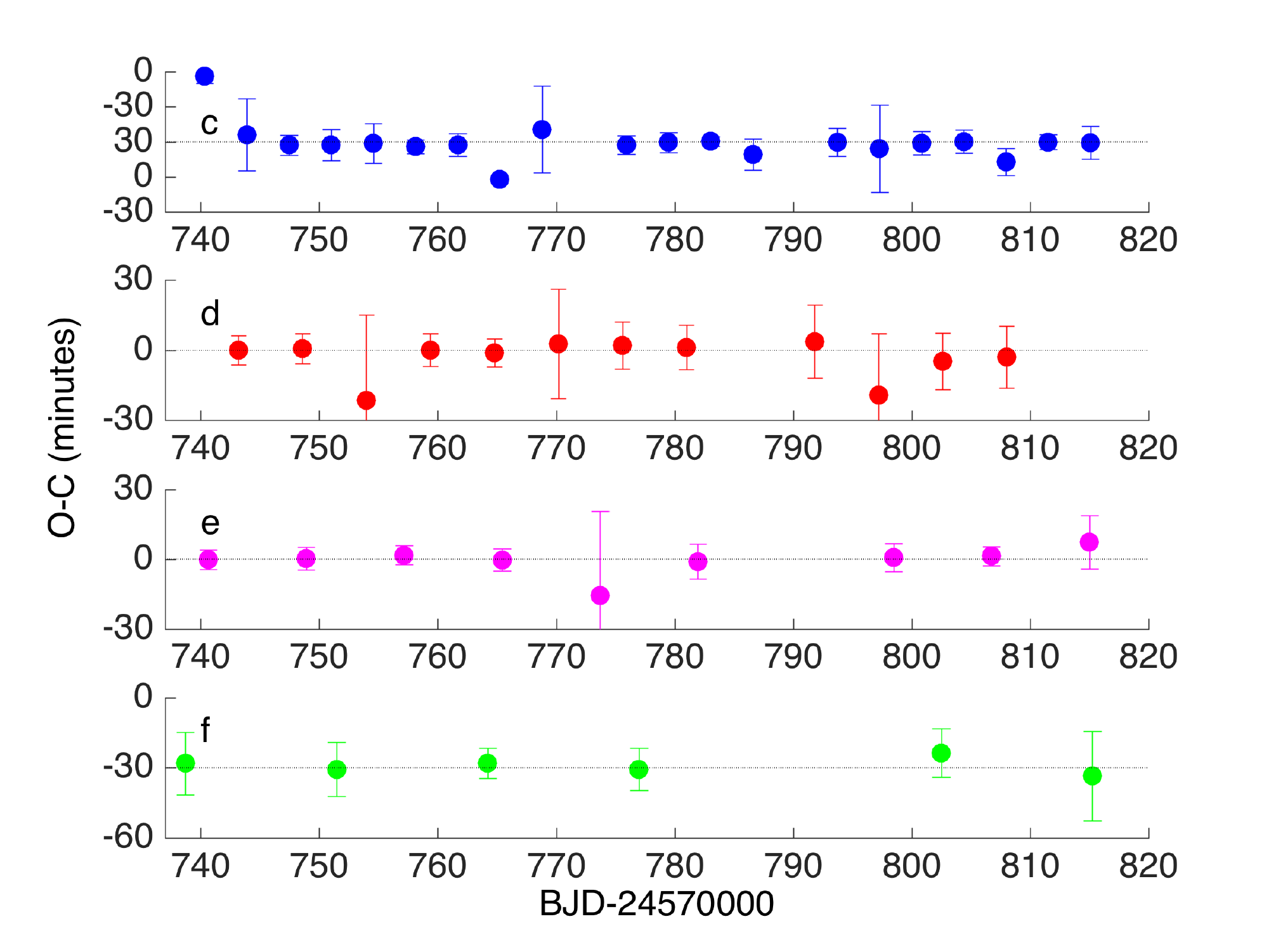}
\caption{Examining the transit times of K2-138 c, d, e, and f. There are no significant variations observed at the timing precision of the \emph{K2} 30-minute cadence observations. \label{fig:ttvs}}
\end{figure}

We calculated the transit times of \ky\ c, d, e, and f shown in Figure \ref{fig:ttvs}. For each transit, we fix the model transit parameters to the best-fit values given in Table \ref{tab:planets}, and allow only the mid-transit time to vary. 
To calculate the uncertainties, we compute the residuals from the best-fit model and perform a bootstrap analysis using the closest 100 timestamps, re-fitting the mid-transit time at each timestamp permutation. \citep{Wall2003}. Examining the resulting transit times, we do not find evidence of significant variations at the level of the average 8--10 minute timing precision from the \emph{K2} data.
The individual transits of \ky~b have insufficient signal-to-noise for robust transit time calculation.

In order to estimate the amplitude of potential transit timing variations (TTVs), we use the mass-radius relation of \citet{Weiss2014} for planet radii in the range 1.5--4R$_{\oplus}$ ($M_p=2.69\times R_p^{0.93}$), predicting that the five planets have masses between 4--7$M_{\oplus}$. Near resonance, TTV amplitudes depend on planet masses, proximity to resonance, and orbital eccentricities. Using the \texttt{TTVFaster} code\footnote{\url{https://github.com/ericagol/TTVFaster}} \citep{Agol2016}, we estimate potential TTV amplitudes of 2.5, 5.1, 7.1, 6.9 and 4.8 minutes for planets b, c, d, e, and f respectively, assuming circular orbits; for eccentric orbits, these amplitudes could be higher. We can also estimate the `super-period' of the planets: when two planets are close to resonance, their transit timing variations evolve on a larger timescale referred to as the super-period. Using Equation 5 from \citep{Lithwick2012b}, we calculate super-periods of 139.4~d, 148.1~d, 144.7~d and 144.2~d for the b-c, c-d, d-e, and e-f pairs, respectively. The \emph{K2} observations span slightly more than half of this amount of time, but as shown the uncertainties on the measured transit times with the processed photometry are large enough to swamp the amplitude of the expected signal. However, with careful sampling over a longer observing baseline and higher precision photometry, the TTVs may be accessible. One possibility is the NASA \emph{Spitzer} telescope. For K2-18b, \citet{Benneke2017} measure a transit timing precision of $\sim$0.9 minutes with \emph{Spitzer}. Using the error approximation of \citet{Carter2008} and scaling for the properties of the \ky\ planets, we estimate that \emph{Spitzer} would achieve transit timing precision of $\sim$2 minutes, which would be sufficient to measure the TTVs of the outer planets. Another possibility for measuring TTVs is the ESA CHEOPS mission \citep{Broeg2013}, although K2-138 ($V=12.2$, $K=10.3$) is at the faint end of their target range.

The empirical relation of \citet{Weiss2014} disguises a large scatter in the measured masses for planets ranging from 2--3R$_{\oplus}$, spanning nearly an order of magnitude from roughly 2--20$M_{\oplus}$ \citep[see Fig.~11 of][]{Christiansen2017}. This diversity is due to a wide, degenerate mix of rock, volatile and gas compositions that can comprise this size of planet. Although the \ky\ planets do not demonstrate significant transit timing variations in the \emph{K2} data, their masses may be accessible to radial velocity observations. By comparing to the ensemble of mass-radius measurements to date, we estimate a minimum mass of $4 M_\oplus$ for the four outer planets, and therefore RV semi-amplitudes of $\gtrsim$2~m~s$^{-1}$. Achieving this precision on \ky\ may be a challenge, given the aforementioned stellar activity level. If any of the planets are measured to be lower density, and therefore likely volatile rich (such as the resonant planets in Kepler-79), they may be interesting yet challenging prospects for atmosphere characterization, given the moderate brightness of the host star.

\section{Conclusions}

We have presented K2-138, the first discovery from the citizen scientists participating in the Exoplanet Explorers project. K2-138 is a compact system of five sub-Neptune-sized planets orbiting an early K star in a chain of successive near-first order resonances; in addition the planets are locked in a set of three-body Laplace resonances. The planets may be accessible to mass measurement via dedicated radial velocity monitoring, and possibly via transit timing variations with improved timing precision. The Exoplanet Explorers project has provided an additional 68 candidate planets from the C12 light curves which likely contain additional planet discoveries, and we plan to upload potential detections for consideration from the rest of the available \emph{K2} campaigns.

\acknowledgments

We thank the anonymous referee for thoughtful and detailed comments that improved the analysis presented in this paper.

This project has been made possible by the contributions of approximately 14,000 volunteers in the Exoplanet Explorers project. The contributions of those volunteers who registered on the project are individually acknowledged at {\url{https://www.zooniverse.org/projects/ianc2/exoplanet-explorers/about/team}}. 

We thank production teams from the British and Australian Broadcasting Corporations as well as Fremantle Media for their help in including Exoplanet Explorers on the \textit{Stargazing Live} programs broadcast in April 2017. 

This paper includes data collected by the \emph{K2} mission. Funding for the \emph{K2} mission is provided by the NASA Science Mission directorate.

This publication uses data generated via the \url{Zooniverse.org} platform, development of which is funded by generous support, including a Global Impact Award from Google, and by a grant from the Alfred P. Sloan Foundation.

BDS acknowledges support from the National Aeronautics and Space Administration (NASA) through Einstein Postdoctoral Fellowship Award Number PF5-160143 issued by the Chandra X-ray Observatory Center, which is operated by the Smithsonian Astrophysical Observatory for and on behalf of NASA under contract NAS8-03060.

This research has made use of the NASA/IPAC Infrared Science Archive, which is operated by the Jet Propulsion Laboratory, California Institute of Technology, under contract with the National Aeronautics and Space Administration. This research has also made use of the NASA Exoplanet Archive and the Exoplanet Follow-up Observation Program website, which are operated by the California Institute of Technology, under contract with the National Aeronautics and Space Administration under the Exoplanet Exploration Program.

Funding for the SDSS and SDSS-II has been provided by the Alfred P. Sloan Foundation, the Participating Institutions, the National Science Foundation, the U.S. Department of Energy, the National Aeronautics and Space Administration, the Japanese Monbukagakusho, the Max Planck Society, and the Higher Education Funding Council for England. The SDSS Web Site is http://www.sdss.org/.



\facility{Kepler, Keck:I (HIRES), Gemini:South (NIRI), Exoplanet Archive, IRSA, MAST, Sloan}



\clearpage

\begin{table*}[h]
\small
\begin{center}
\caption{Final \ky\  system planet parameters from the five-planet transit model. \label{tab:planets}}
\resizebox{\textwidth}{!}{%
\begin{tabular}{llllll}
\hline
Parameter & b & c & d & e & f\\
\hline
Period (d) & 
$2.35322\pm0.00036$ & 
$3.55987^{+0.00023}_{-0.00022}$ &
$5.40478^{+0.00048}_{-0.00046}$ &
$8.26144^{+0.00045}_{-0.00044}$ &
$12.75759^{+0.00092}_{-0.00092}$ \\
T$_0$ (BJD) &
$2457773.3170^{+0.0037}_{-0.0038}$ &
$2457740.3223^{+0.0025}_{-0.0027}$ &
$2457743.1607^{+0.0036}_{-0.0037}$ &
$2457740.6451^{+0.0020}_{-0.0021}$ &
$2457738.7019^{+0.0033}_{-0.0035}$ \\
T$_{14}$ (hr) &
$1.73^{+0.23}_{-0.23}$ & 
$2.329^{+0.095}_{-0.088}$ & 
$2.97^{+0.13}_{-0.11}$ &  
$3.063^{+0.107}_{-0.085}$ & 
$3.19^{+0.14}_{-0.12}$ \\
R$_p$/R$_\star$ &
$0.0168^{+0.0029}_{-0.0017}$ &
$0.0267^{+0.0036}_{-0.0015}$ &
$0.0283^{+0.0041}_{-0.0017}$ &
$0.0349^{+0.0037}_{-0.0015}$ &
$0.0299^{+0.0039}_{-0.0018}$ \\
R$_\star$/a & 
$0.113^{+0.052}_{-0.021}$ &
$0.0930^{+0.0361}_{-0.0097}$ &
$0.0784^{+0.0307}_{-0.0085}$ &
$0.0516^{+0.0176}_{-0.0047}$ &
$0.0356^{+0.0136}_{-0.0040}$ \\
a (AU) & 
$0.03380^{+0.00024}_{-0.00024}$ &
$0.04454^{+0.00032}_{-0.00032}$ &
$0.05883^{+0.00042}_{-0.00042}$ &
$0.07807\pm0.00056$ &
$0.10430^{+0.00074}_{-0.00075}$ \\
i (deg) &
$86.9^{+2.2}_{-4.6}$ &       
$87.5^{+1.8}_{-3.3}$ &       
$87.9^{+1.5}_{-2.8}$ &        
$88.70^{+0.93}_{-1.66}$ & 
$89.03^{+0.70}_{-1.22}$ \\ 
b & 
$0.50^{+0.33}_{-0.34}$ &
$0.47\pm0.32$ &
$0.47^{+0.31}_{-0.32}$ &
$0.44^{+0.31}_{-0.30}$ &
$0.48^{+0.30}_{-0.33}$ \\
e (1-$\sigma$ upper limit) & 
$<$0.403 & 
$<$0.296 & 
$<$0.348 & 
$<$0.315 & 
$<$0.364 \\
R$_p$ (R$_{\oplus}$) & 
$1.57^{+0.28}_{-0.17}$ &
$2.52^{+0.34}_{-0.16}$ &
$2.66^{+0.39}_{-0.18}$ &
$3.29^{+0.35}_{-0.18}$ &
$2.81^{+0.36}_{-0.19}$ \\
Insolation ($I_{\oplus}$) & 
$486^{+37}_{-35}$ & 
$279^{+21}_{-20}$ & 
$160^{+12}_{-11}$ & 
$91.1^{+7.0}_{-6.6}$ & 
$51.0^{+3.9}_{-3.7}$ \\
User votes & 
- &
12/12 (100\%) & 
14/14 (100\%) & 
15/15 (100\%) & 
12/13 (93\%) \\
\tableline
\end{tabular}}
\end{center}
\end{table*}

%

\clearpage


\begin{thebibliography}{}
\bibliographystyle{apj}
\bibitem[Agol \& Deck(2016)]{Agol2016} Agol, E., \& Deck, K.\ 2016, \apj, 818, 177 
\bibitem[Astudillo-Defru et al.(2015)]{Astudillo2015} Astudillo-Defru, N., Bonfils, X., Delfosse, X., et al.\ 2015, \aap, 575, A119 
\bibitem[Astudillo-Defru et al.(2017)]{Astudillo2017} Astudillo-Defru, N., Forveille, T., Bonfils, X., et al.\ 2017, arXiv:1703.05386 
\bibitem[Barclay et al.(2015)]{Barclay2015} Barclay, T., Quintana, E.~V., Adams, F.~C., et al.\ 2015, \apj, 809, 7 
\bibitem[Barentsen (2017)]{Barentsen2017} Barentsen, G., Keplergo/kadenza: v2.0.2 (2017). URL https://doi.org/10.5281/zenodo.344973.
\bibitem[Batygin \& Morbidelli(2013)]{Batygin2013} Batygin, K., \& Morbidelli, A.\ 2013, \aj, 145, 1 
\bibitem[Benneke et al.(2017)]{Benneke2017} Benneke, B., Werner, M., Petigura, E., et al.\ 2017, \apj, 834, 187 
\bibitem[Broeg et al.(2013)]{Broeg2013} Broeg, C., Fortier, A., Ehrenreich, D., et al.\ 2013, European Physical Journal Web of Conferences, 47, 03005 
\bibitem[Cabrera et al.(2017)]{Cabrera2017} Cabrera, J., Barros, S.~C.~C., Armstrong, D., et al.\ 2017, arXiv:1707.08007 
\bibitem[Carter et al.(2008)]{Carter2008} Carter, J.~A., Yee, J.~C., Eastman, J., Gaudi, B.~S., \& Winn, J.~N.\ 2008, \apj, 689, 499-512 
\bibitem[Casey et al.(2016)]{Casey2016} Casey, A.~R., Hawking, K., Hogg, D.~W., et al.,\ 2016, arXiv:1609.02914
\bibitem[Christiansen et al.(2017)]{Christiansen2017} Christiansen, J.~L., Vanderburg, A., Burt, J., et al.\ 2017, \apj, in press 
\bibitem[Ciardi et al.(2015)]{Ciardi2015} Ciardi, D.~R., Beichman, C.~A., Horch, E.~P., \& Howell, S.~B.\ 2015, \apj, 805, 16 
\bibitem[Crossfield et al.(2015)]{Crossfield2015} Crossfield, I.~J.~M., Petigura, E., Schlieder, J.~E., et al.\ 2015, \apj, 804, 10 
\bibitem[Crossfield et al.(2016)]{Crossfield2016} Crossfield, I.~J.~M., Ciardi, D.~R., Petigura, E.~A., et al.\ 2016, \apjs, 226, 7 
\bibitem[Crossfield et al.(2017)]{Crossfield2017} Crossfield, I.~J.~M., Ciardi, D.~R., Isaacson, H., et al.\ 2017, arXiv:1701.03811 
\bibitem[Dai et al.(2016)]{Dai2016} Dai, F., Winn, J.~N., Albrecht, S., et al.\ 2016, \apj, 823, 115 
\bibitem[Dotter et al.(2008)]{Dotter2008} Dotter, A., Chaboyer, B., Jevremovi{\'c}, D., et al.\ 2008, \apjs, 178, 89-101 
\bibitem[Eastman et al.(2013)]{Eastman2013} Eastman, J., Gaudi, B.~S., \& Agol, E.\ 2013, \pasp, 125, 83 
\bibitem[Fabrycky et al.(2014)]{Fabrycky2014} Fabrycky, D.~C., Lissauer, J.~J., Ragozzine, D., et al.\ 2014, \apj, 790, 146  
\bibitem[Fischer et al.(2012)]{Fischer2012} Fischer, D.~A., Schwamb, M.~E., Schawinski, K., et al.\ 2012, \mnras, 419, 2900 
\bibitem[Ford et al.(2012)]{Ford2012} Ford, E.~B., Ragozzine, D., Rowe, J.~F., et al.\ 2012, \apj, 756, 185 
\bibitem[Foreman-Mackey et al.(2013)]{Foreman2013} Foreman-Mackey, D., Hogg, D.~W., Lang, D., \& Goodman, J.\ 2013, \pasp, 125, 306 
\bibitem[Furlan et al.(2017)]{Furlan2017} Furlan, E., Ciardi, D.~R., Everett, M.~E., et al.\ 2017, \aj, 153, 71 
\bibitem[Gillon et al.(2016)]{Gillon2016} Gillon, M., Jehin, E., Lederer, S.~M., et al.\ 2016, \nat, 533, 221 
\bibitem[Gillon et al.(2017)]{Gillon2017} Gillon, M., Triaud, A.~H.~M.~J., Demory, B.-O., et al.\ 2017, \nat, 542, 456 
\bibitem[Gladman(1993)]{Gladman1993} Gladman, B.\ 1993, \icarus, 106, 247 
\bibitem[Go{\'z}dziewski et al.(2016)]{Gozd2016} Go{\'z}dziewski, K., Migaszewski, C., Panichi, F., \& Szuszkiewicz, E.\ 2016, \mnras, 455, L104 
\bibitem[Hodapp et al.(2003)]{Hodapp2003} Hodapp, K.~W., Jensen, J.~B., Irwin, E.~M., et al.\ 2003, \pasp, 115, 1388 
\bibitem[Howard et al.(2012)]{Howard2012} Howard, A.~W., Marcy, G.~W., Bryson, S.~T., et al.\ 2012, \apjs, 201, 15 
\bibitem[Howell et al.(2014)]{Howell2014} Howell, S.~B., Sobeck, C., Haas, M., et al.\ 2014, \pasp, 126, 398 
\bibitem[Huber et al. (2016)]{Huber2016} Huber, D., Bryson, S.~T., Haas, M.~R., et al.\ 2016, \apjs, 224, 2
\bibitem[Johnson et al.(2017)]{Johnson2017} Johnson, J.~A., Petigura, E.~A., Fulton, B.~J., et al.\ 2017, arXiv:1703.10402
\bibitem[Jontof-Hutter et al.(2014)]{Jontof-Hutter2014} Jontof-Hutter, D., Lissauer, J.~J., Rowe, J.~F., \& Fabrycky, D.~C.\ 2014, \apj, 785, 15 \bibitem[Kipping(2013)]{Kipping2013} Kipping, D.~M.\ 2013, \mnras, 434, L51 
\bibitem[Kolbl et al.(2015)]{Kolbl2015} Kolbl, R., Marcy, G.~W., Isaacson, H., \& Howard, A.~W.\ 2015, \aj, 149, 18 
\bibitem[Kunder et al.(2017)]{Kunder2017} Kunder, A., Kordopatis, G., Steinmetz, M., et al.\ 2017, \aj, 153, 75
\bibitem[Latham et al.(2011)]{Latham2011} Latham, D.~W., Rowe, J.~F., Quinn, S.~N., et al.\ 2011, \apjl, 732, L24 
\bibitem[Lillibridge et al.(2001)]{CAPCHA} Lillibridge, M. D., Abadi, M., Bharat, K. \& Broder, A. Z. 2001, US Patent 6,195,698. Washington, DC: U.S.
\bibitem[Lintott et al.(2008)]{Lintott2008} Lintott, C.~J., Schawinski, K., Slosar, A., et al.\ 2008, \mnras, 389, 1179
\bibitem[Lissauer et al.(2011)]{Lissauer2011} Lissauer, J.~J., Ragozzine, D., Fabrycky, D.~C., et al.\ 2011, \apjs, 197, 8 
\bibitem[Lissauer et al.(2012)]{Lissauer2012} Lissauer, J.~J., Marcy, G.~W., Rowe, J.~F., et al.\ 2012, \apj, 750, 112
\bibitem[Lithwick \& Wu(2012)]{Lithwick2012} Lithwick, Y., \& Wu, Y.\ 2012, \apjl, 756, L11 
\bibitem[Lithwick et al.(2012)]{Lithwick2012b} Lithwick, Y., Xie, J., \& Wu, Y.\ 2012, \apj, 761, 122 
\bibitem[Luger et al.(2017)]{Luger2017} Luger, R., Sestovic, M., Kruse, E., et al.\ 2017, Nature Astronomy, 1, 0129 
\bibitem[Mandel \& Agol(2002)]{Mandel2002} Mandel, K., \& Agol, E.\ 2002, \apjl, 580, L171 \bibitem[MacDonald et al.(2016)]{MacDonald2016} MacDonald, M.~G., Ragozzine, D., Fabrycky, D.~C., et al.\ 2016, \aj, 152, 105 
\bibitem[Mills et al.(2016)]{Mills2016} Mills, S.~M., Fabrycky, D.~C., Migaszewski, C., et al.\ 2016, \nat, 533, 509 
\bibitem[Morton(2012)]{Morton2012} Morton, T.~D.\ 2012, \apj, 761, 6
\bibitem[Morton(2015)]{Morton2015} Morton, T.~D.\ 2015, isochrones: Stellar model grid package,
Astrophysics Source Code Library 
\bibitem[Morton et al.(2016)]{Morton2016} Morton, T.~D., Bryson, S.~T., Coughlin, J.~L., et al.\ 2016, \apj, 822, 86 
\bibitem[Pecaut \& Mamajek (2013)]{Pecaut2013} Pecaut, M.~J. \& Mamjek, E.~E.\ 2013, \apjs, 208, 9
\bibitem[Petigura et al.(2013a)]{Petigura2013a} Petigura, E.~A., Marcy, G.~W., \& Howard, A.~W.\ 2013a, \apj, 770, 69
\bibitem[Petigura et al.(2013b)]{Petigura2013b} Petigura, E.~A., Howard, A.~W., \& Marcy, G.~W.\ 2013b, Proceedings of the National Academy of Science, 110, 19273 
\bibitem[Petigura et al.(2015)]{Petigura2015} Petigura, E.~A., Schlieder, J.~E., Crossfield, I.~J.~M., et al.\ 2015, \apj, 811, 102 
\bibitem[Petigura et al.(2017)]{Petigura2017} E.~A., Howard, A.~W., Marcy, G.~W. et al.\ 2017, arXiv:1703.10400
\bibitem[Ramos et al.(2017)]{Ramos2017} Ramos, X.~S., Charalambous, C., Ben{\'{\i}}tez-Llambay, P., \& Beaug{\'e}, C.\ 2017, arXiv:1704.06459 
\bibitem[Rasmussen and Williams(2005)]{Rasmussen2005} Rasmussen C. and Williams C. 2005 \emph{{G}aussian Processes for Machine Learning (Adaptive Computation and Machine Learning)} The MIT Press, Cambridge, MA, USA, 2005.
\bibitem[Rodriguez et al.(2017)]{Rodriguez2017} Rodriguez, J.~E., Zhou, G., Vanderburg, A., et al.\ 2017, arXiv:1701.03807 
\bibitem[Rowe et al.(2014)]{Rowe2014} Rowe, J.~F., Bryson, S.~T., Marcy, G.~W., et al.\ 2014, \apj, 784, 45 
\bibitem[Rowe et al.(2015)]{Rowe2015} Rowe, J.~F., Coughlin, J.~L., Antoci, V., et al.\ 2015, \apjs, 217, 16 
\bibitem[Siebert et al.(2011)]{Siebert2011} Siebert, A., Williams, M.~.E.~K., Siviero, A., et al.\ 2011, \aj, 141, 187
\bibitem[Smalley et al.(2012)]{Smalley2012} Smalley, B., Anderson, D.~R., Collier-Cameron, A., et al.\ 2012, \aap, 547, A61 
\bibitem[Sinukoff et al.(2016)]{Sinukoff2016} Sinukoff, E., Howard, A.~W., Petigura, E.~A., et al.\ 2016, \apj, 827, 78 
\bibitem[Steinmetz et al.(2006)]{Steinmetz2006} Steinmetz, M., Zwitter, T., Siebert, A., et al.\ 2006, \aj, 132, 1645
\bibitem[Vanderburg et al.(2016)]{Vanderburg2016} Vanderburg, A., Bieryla, A., Duev, D.~A., et al.\ 2016, \apjl, 829, L9 
\bibitem[von Ahn et al.(2003)]{vonAhn2003} von Ahn, L., Blum, M., Hopper, N. J., Langford, J. 2003. CAPTCHA: Using Hard AI Problems for Security. EUROCRYPT 2003: International Conference on the Theory and Applications of Cryptographic Techniques.
\bibitem[Wall et al.(2003)]{Wall2003} Wall, J.~V., Jenkins, C.~R., Ellis, R., et al.\ 2003, Practical statistics for astronomers (Cambridge University Press)
\bibitem[Wang et al.(2017)]{Wang2017} Wang, S., Wu, D.-H., Barclay, T., \& Laughlin, G.~P.\ 2017, arXiv:1704.04290 
\bibitem[Watson et al.(2000)]{Watson2000} Watson, F.~G., Parker, Q.~A., Bogatu, G., et al.\ 2000, ProcSPIE, 4008, 123
\bibitem[Weiss \& Marcy(2014)]{Weiss2014} Weiss, L.~M., \& Marcy, G.~W.\ 2014, \apjl, 783, L6 
\bibitem[Winn \& Fabrycky(2015)]{Winn2015} Winn, J.~N., \& Fabrycky, D.~C.\ 2015, \araa, 53, 409 
\bibitem[Xie(2013)]{Xie2013} Xie, J.-W.\ 2013, \apjs, 208, 22 
\bibitem[Xie(2014)]{Xie2014} Xie, J.-W.\ 2014, \apjs, 210, 25 



\end{thebibliography}
\end{document}